\documentclass[10pt,conference]{IEEEtran}
\IEEEoverridecommandlockouts

\usepackage{cite}
\usepackage{amsmath,amssymb,amsfonts}
\usepackage{algorithmic}
\usepackage{graphicx}
\usepackage{textcomp}
\usepackage{xcolor}
\usepackage{booktabs}
\usepackage{multirow}
\usepackage{xspace}
\usepackage{array}
\usepackage[flushleft]{ragged2e}
\usepackage{makecell}
\usepackage{soul}
\usepackage{color}
\usepackage{colortbl}
\usepackage{balance}

\newcommand{\mycomment}[2]{\{\textbf{#1:} #2\}}
\newcommand{\tool}{\textsc{ReuseDroid}\xspace}
\newcommand{\lin}{\textit{Lin}\xspace}
\newcommand{\dataset}{\textit{LinPro}\xspace}
\newcommand{\numMig}{578\xspace}
\newcommand{\taa}{Test Analyzer Agent\xspace}
\newcommand{\pa}{Planner Agent\xspace}
\newcommand{\ea}{Execution Agent\xspace}
\newcommand{\fa}{Feedback Agent\xspace}

\newcommand{\xiaolei}[1]{\textcolor{blue}{\mycomment{xiaolei}{#1}}}
\newcommand{\bella}[1]{\textcolor{green}{\mycomment{bella}{#1}}}
\newcommand{\yepang}[1]{\textcolor{purple}{\mycomment{yepang}{#1}}}
\newcommand{\scc}[1]{\textcolor{red}{\mycomment{scc}{#1}}}

\renewcommand{\xiaolei}[1]{}
\renewcommand{\bella}[1]{}
\renewcommand{\yepang}[1]{}
\renewcommand{\scc}[1]{}

\def\BibTeX{{\rm B\kern-.05em{\sc i\kern-.025em b}\kern-.08em
    T\kern-.1667em\lower.7ex\hbox{E}\kern-.125emX}}
\begin{document}

\title{ReuseDroid: A VLM-empowered Android UI Test Migrator Boosted by Active Feedback}

\author{
\IEEEauthorblockN{Xiaolei Li\IEEEauthorrefmark{1}\IEEEauthorrefmark{2},
Jialun Cao\IEEEauthorrefmark{1},
Yepang Liu\IEEEauthorrefmark{2},
Shing-Chi Cheung\IEEEauthorrefmark{1},
Hailong Wang\IEEEauthorrefmark{2}
}
\text{\small \{xlihx, jcaoap, scc\}@cse.ust.hk, liuyp1@sustech.edu.cn, 12112319@mail.sustech.edu.cn}
\IEEEauthorblockA{\IEEEauthorrefmark{1}Department of Computer Science and Engineering, The Hong Kong University of Science and Technology, China}
\IEEEauthorblockA{\IEEEauthorrefmark{2}Department of Computer Science and Engineering, Southern University of Science and Technology, China}
}

\maketitle

\begin{abstract}
  GUI testing is an essential quality assurance process in mobile app development. However, the creation and maintenance of GUI tests for mobile apps are resource-intensive and costly.
  Recognizing that many apps share similar functionalities, researchers have proposed various techniques to migrate GUI tests from one app to another with similar features.
  For example, some techniques employ mapping-based approaches to align the GUI elements traversed by the tests of a source app to those present in the target app.
  Other test migration techniques have also been proposed to leverage large language models (LLMs) by adapting the GUI tasks in source tests.
  However, these techniques are ineffective in dealing with different operational logic between the source and target apps. The semantics of GUI elements may not be correctly inferred due to the missing analysis of these flows.
  In this work, we propose \textbf{\tool}, a novel multi-agent framework for GUI test migration empowered by Large Vision-Language Models (VLMs).
  \tool is powered by multiple VLM-based agents,
  each tackling a stage of the test migration process by leveraging the relevant visual and textual information embedded in GUI pages.
  An insight of \tool is to migrate tests based only on the core logic shared across similar apps, while their entire operational logic could differ.
  We evaluate \tool on \dataset, a new test migration dataset that consists of \numMig migration tasks for 39 popular apps across 4 categories.
  The experimental result shows that \tool can successfully migrate 90.3\% of the migration tasks, 
  outperforming the best mapping-based and LLM-based baselines by 318.1\% and 109.1\%, respectively.
\end{abstract}


\maketitle

\section{Introduction}\label{sec:introduction}

\begin{figure*}
    \centering
    \includegraphics[width=0.9\textwidth]{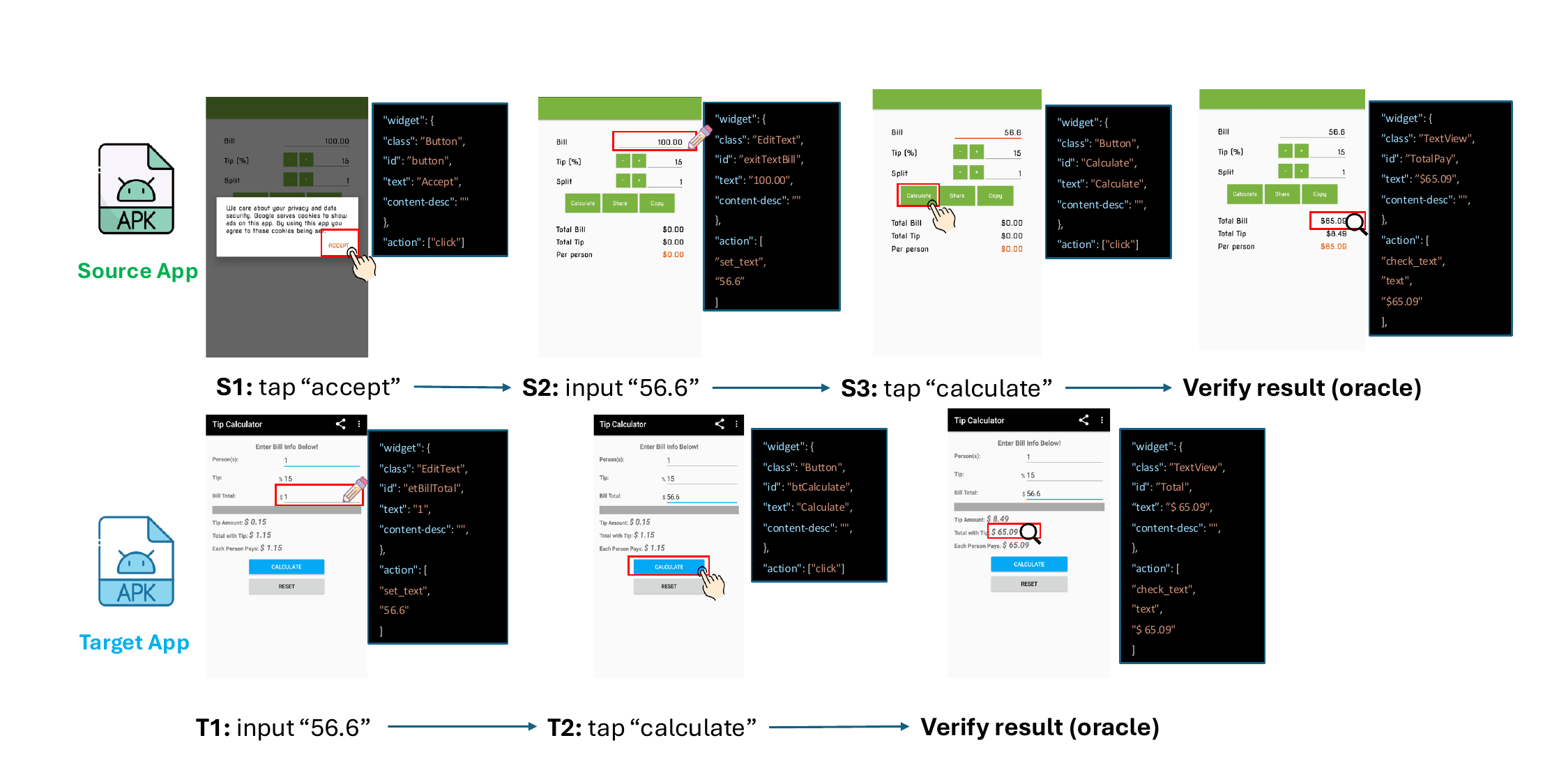}
    \caption{An example of GUI test migration}
    \label{fig:intro_example}
\end{figure*}

GUI testing is a crucial process in mobile application development. It helps ensure the usability and stability of the business logic embedded in an application. 
To ease the test construction effort, GUI test migration has been proposed \cite{stoat2017,atm2019}.
An example of test migration for the calculator function is shown in Figure~\ref{fig:intro_example}.
The key to test migration lies in the similarity of operational logic that supports the same functionality between two applications.
During the test migration, the operations in the source test are mapped to the corresponding widgets in the target app.

\par
Existing techniques \cite{craftdroid2019, temdroid2024, migratepro2024} utilize an action-by-action mapping framework, where each operation in the source test is mapped to a corresponding widget in the target app.
The process continues iteratively until all operations in the source test are mapped to target widgets, forming a target test case.
A key challenge arises because of the widget-level differences between the source and target apps. Specifically, the identifiers of the widgets in the source app may differ from those in the target app.
In the example, the \textit{id} of the bill amount input widget in the source app is different from that in the target app, making it challenging to find the corresponding target widget for the operation \textbf{S3}.
To address this challenge, prior works employ language embedding models, like word2vec\cite{word2vec} and BERT\cite{bert}, to compute the semantic similarity between widgets.
By mapping the source operations to the most semantically similar target widgets, they effectively bridge the gap at this level. 
However, more unresolved challenges remain at the interaction level differences, illustrated in the example in Figure~\ref{fig:motivation_example}.
To illustrate the challenges, we define the following terms: operation/action/event and \textbf{logic step}.
In this paper, we use operation, action, and event to refer to atomic operations and logic step to refer to a subtask that contributes to the completion of the target functionality.
A logic step could be achieved by one or multiple operations.
These challenges can be categorized into three key aspects:

\textbf{a.} The operational logic that triggers the target functionality typically consists of multiple logic steps. A logic step is a subtask that contributes to the completion of the target functionality.
Due to differences in design philosophies, the number and sequence of logic steps in the source app may differ from those in the target app.
For example, the source app in Figure~\ref{fig:motivation_example} does not require a logic step to confirm the bill amount by tapping a button, while the target app does.
\textbf{b.} Even for the same logic step, the required sequence of operations to complete it may vary.
For example, the source app requires four tapping operations to input a bill amount, while the target app requires one inputting operation.
\textbf{c.} Even if the operation sequence is identical, the specific actions needed to activate the corresponding UI elements may differ, like a button that requires a single tap in the source app but a long press in the target app.
With the action-by-action mapping framework, existing approaches prove to be ineffective in handling these challenges, 
as they always try to map each source action to a target widget, while a one-to-one mapping is not always available\cite{sail2024}.

\par
To explore solutions to these challenges, we adapt LLM-based GUI task automation methods\cite{rebl2024,autodroid2024} for GUI test migration. 
These methods adopt an exploration-based framework in which the LLMs are prompted to extract useful information from the source test.
Subsequently, this information will be used as guidance to explore the target app's GUI pages, trigger the target functionality, and generate the interaction sequences accordingly.
The exploration process stops when the LLMs determine that the target functionality has been successfully triggered.
These methods have shown promise in GUI test migration, as they show better performance in understanding the high-level intent of the source test and the GUI pages of the target app.
Also, the exploration-based framework is more flexible since it does not rely on the one-to-one mapping of source actions to target widgets but rather focuses on finding appropriate interactions to meet the high-level intent of the source test.
Specifically, they are effective in understanding the logic steps and inferring the correct interaction sequences to complete each logic step.
However, several challenges persist:

\textbf{1)} When the source test contain redundant operations that are not necessary for the target app, existing exploration-based methods fail to identify and eliminate them, causing exploration to be misled or get stuck.
\textbf{2)} The ambiguity in the source test, caused by the meaningless, unclear, or empty descriptions in the source test, cannot be identified by existing methods, thus leading to misinterpretation of the source test's intent.
\textbf{3)} The complexity of the target app's GUI pages may cause the LLMs to hallucinate, generating incorrect actions that do not trigger the target functionality.
\textbf{4)} The methods show poor performance in recognizing the stopping conditions, leading to either premature termination or endless exploration.

\par
To address these challenges, we propose a multi-agent framework for GUI test migration based on Large Vision-Language Models (VLMs) called \tool.
Inspired by LLM-based GUI automation tools,   \tool employs an exploration-based framework to dynamically explore the target app and generate the target test.
We introduce a \taa to generalize logic steps from atomic operations in the source test and eliminate redundant logic steps.
The key insight here is that the key logic underlying the target functionality is universally shared across applications, 
while the ones handling app-specific behaviors could be not, which we refer to as supporting steps.
In Figure~\ref{fig:motivation_example}, the key logic is to provide the bill amount as source data, and the supporting steps are the steps handling user tutorials pages, input confirmation(tap calculate button), etc.
Therefore, we identify the steps carrying the key logic and preserve them while eliminating the supporting steps to prevent misleading the exploration process, addressing Challenge 1.

In the app exploration phase, we introduce a dedicated Completeness Checking module to explicitly determine whether the target functionality has been successfully triggered, addressing Challenge 4.

\tool may generate incorrect actions in the exploration process because of complex GUI pages or hallucinations.
In other words, \tool mispredicts the consequences of the action when generating it.
We introduce a \fa that watches the execution of each generated action and provides feedback to correct the mistakes.
\fa has reduced the hallucination problem introduced by the complex GUI pages and the VLM itself, addressing Challenge 3.

All these modules utilize Vision-Language Models (VLMs) that utilize screenshots from the source app and target app as supplementary contexts.
The screenshots provide additional information about the source test and the target app, which is essential for alleviating the semantic insufficiency and ambiguity in the source test and the target app, addressing Challenges 2 and 3.

In addition, we expand an existing dataset \lin\cite{craftdroid2019} to improve its scale and representativeness and evaluate \tool on this dataset.
The expanded dataset is called \dataset.
\par
To conclude, we propose these contributions in this paper:
\begin{itemize}
    \item We propose a VLM-based multi-agent framework for GUI test migration, called \tool, to address the challenges of operational logic differences in GUI test migration.
    \item We construct a new dataset, \dataset, to evaluate \tool in a more scaled and representative manner.
    \item We evaluate \tool and existing approaches on \dataset, demonstrating \tool's effectiveness and identifying further challenges in GUI test migration.
\end{itemize}
\section{Motivation}\label{sec:motivation}
In this section, we provide motivation examples to illustrate the challenges of GUI test migration.
We select three representative tools and analyze their limitations in handling the challenges.

\subsection{Subject Tools}\label{subsec:subject_tools}
We consider two types of existing tools in this study: mapping-based tools and LLM-based tools. 
For the former, we choose CraftDroid\cite{craftdroid2019} as a representative since it achieved the second-best performance in GUI test migration.
Although TEMDroid\cite{temdroid2024} is a more recent mapping-based tool that achieved state-of-the-art performance,
we do not include it in this study because its public implementation is incomplete.
For the latter, we choose AutoDroid\cite{autodroid2024} and ReBL\cite{rebl2024} as representatives.
AutoDroid generates UI automation scripts according to user instructions on specific UI tasks, while ReBL relies on user-provided bug reports to guide exploration among GUI pages.
Both tools utilize LLMs to extract UI information from GUI pages' DOM trees and understand the intention of GUI pages and widgets.
We choose AutoDroid and ReBL as representatives of LLM-based tools because they are two of the state-of-the-art tools in UI automation, and can easily be adapted to the test migration task.

\subsection{Motivation Example}\label{subsec:motivation_example}
\begin{figure}
    \centering
    \includegraphics[width=0.5\textwidth]{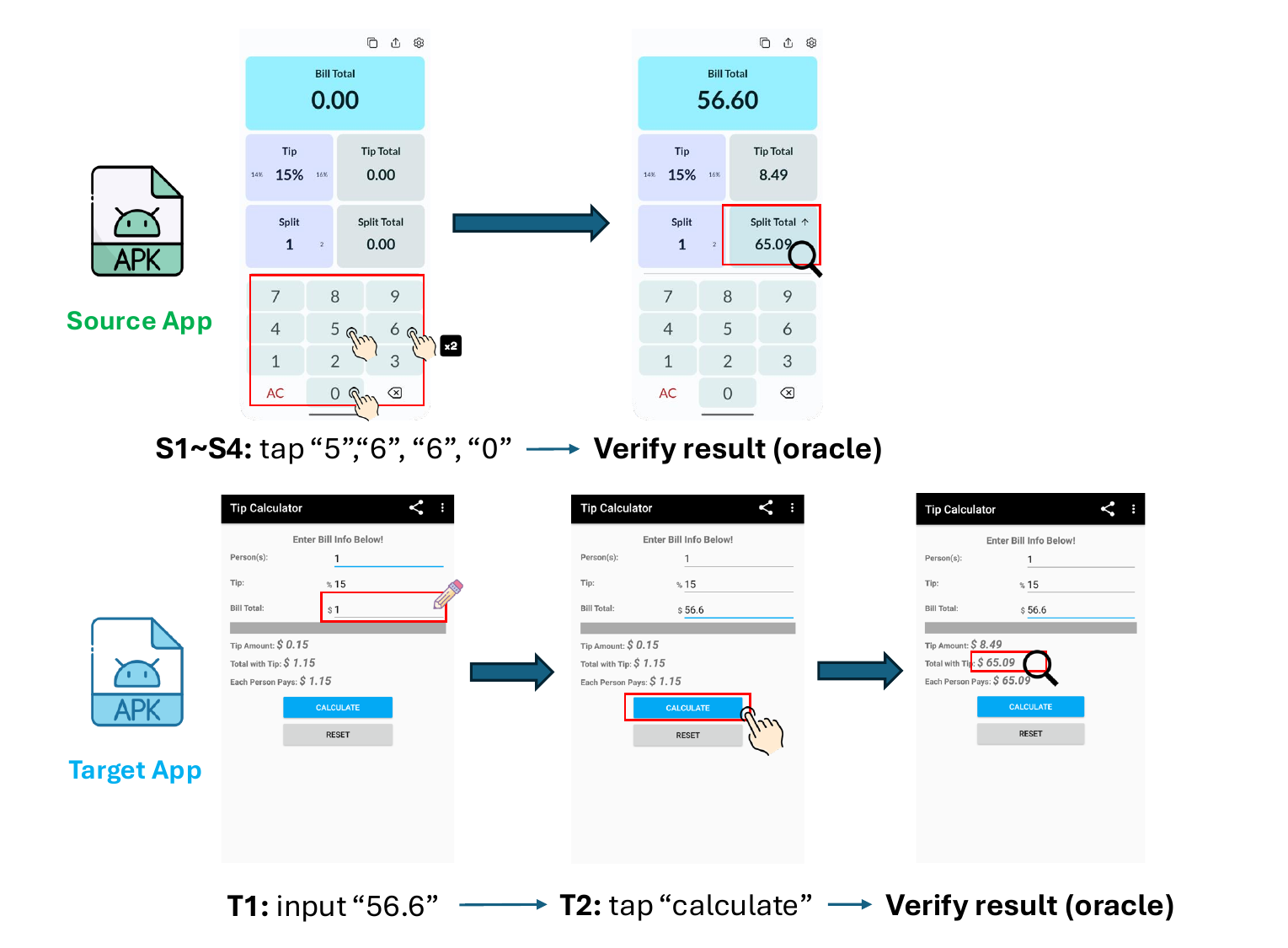}
    \caption{A motivation example}
    \label{fig:motivation_example}
\end{figure}

We examine three subject tools on the \dataset dataset, and present a motivation example to illustrate the challenges remain unsolved by these tools.
The example is shown in Figure~\ref{fig:motivation_example}.
The source test's operational logic consists of four actions: 1) tap '5', 2) tap '6', 3) tap '6', and 4) tap '0'. 
After these actions, the bill amount is set to 56.60, and the total amount is shown as 65.09, verified by the oracle event.
To test the same functionality in the target app, the correct operational logic should be: 1) set the text field bill amount to 56.60, and 2) tap the calculate button.
Then an oracle event is needed to verify the total amount is 65.09.
To conclude, the source test contains one logic step to set the bill amount to 56.60 and an oracle event.
The correct target test contains two logic steps to set the bill amount and calculate the total amount, as well as an oracle event.
It is clear that although the source and target apps share the same functionality, the operational logic to trigger the functionality is different.
Also, the specific interactions to complete the same logic step can be different.

\subsubsection{Limitations of Mapping-based Tools}
Mapping-based frameworks are widely adopted in GUI test migration, and CraftDroid is a representative mapping-based tool.
CraftDroid employs a similarity-based mapping approach to gradually migrate each source action to the target app.
For each source action, CraftDroid calculates the subject source widget's similarity against all widgets in the target app's explored GUI pages.
The widget with the highest similarity is selected as the target widget, and the action on the source widget is mapped to the action on the target widget.
Such mapping-based tools are effective when the migration meets the following conditions:
1) The source and target apps have so similar operational logic to trigger the same functionality that the mapping can be established one-to-one.
2) The actions to interact with the corresponding widgets in the source and target apps are the same.
When the operational logic to trigger the same functionality diverges significantly, the mapping-based tools exhibit limitations.
\par
\textbf{Limitation 1: Ineffective Handling of N-to-n and n-to-N Mappings.} 
It is common that in the operational logic of a certain functionality, some actions collectively achieve a purpose.
In one app, the sequence of actions to achieve the same purpose can be very different from another app.
In Figure~\ref{fig:motivation_example}, four actions are needed in the source app to set the bill amount to 56.60, while only one input action is needed in the target app.
CraftDroid cannot recognize that the four actions in the source app should be considered as a whole during migration but tries to map each action to the target app.
However, when searching for the counterpart for the first action \textit{tap `5'}, there is no such a widget like '5' button in the target app, so CraftDroid gets stuck or chooses a wrong widget to interact with, leading to a failed migration.
In other words, although the source and target test share the same sequence of logic steps, the specific operations to fulfill the logic steps can be different.
In addition, the source and target test may contain different logic steps to trigger the same functionality due to different design philosophies.
Specifically, some apps may include additional logic steps, such as mini-tutorials, hint popups, etc., for purposes like user experience enhancement, while others may simplify the operational logic to reduce user effort.
As a result, the source test may contain redundant logic steps that have no corresponding actions in the target app or miss logic steps that are required in the target app.
For example, in Figure~\ref{fig:motivation_example}, the source app contains an input confirmation step by tapping the `calculate' button, while the target app does not.
Mapping-based tools often show ineffectiveness in handling such differences in operational logic because they simply transfer the source test by searching for the counterpart for each action in the test.
In conclusion, existing mapping-based tools, because of their one-by-one mapping strategy, exhibit limitations in handling the differences in operational logic between the source and target apps,
in both the operation level and logic step level.

\par
\textbf{Limitation 2: Naive Action Migration.}
It is common for the specific operations to activate similar widgets in different apps to be different.
For example, in Figure~\ref{fig:motivation_example}, 
the source app uses a swipe action on the tip percentage slider to set its value,
while the target app uses a set text action.
CraftDroid fails to migrate such actions because it simply executes the source action on the target widget.

\subsubsection{Limitations of LLM-based Tools}
Unlike mapping-based tools, the LLM-based tools studied in this work do not rely on the one-by-one mapping of UI elements between the source and target apps. Instead, they leverage a target-oriented exploration framework to trigger the target functionality.
Specifically, they extract key information from the source test and utilize the information to guide the real-time exploration of the target app's GUI pages, trigger the target functionality, and generate the interaction sequences accordingly.
ReBL, for example, treats the source test as a bug report and the target functionality as the bug to reproduce,
ReBL tries to reproduce the actions in the report to trigger the target functionality in the target app.
Existing LLM-based tools, like AutoDroid and ReBL, although they show promise in general GUI automation tasks and bug report reproduction, exhibit limitations when facing challenges unique to test migration.
\par
\textbf{Limitation 1: Ineffective Handling of Redundant Operations.}

As addressed in Limitation 1 of mapping-based tools, the source test may contain redundant operations and miss necessary operations to trigger the target functionality on the target app.
We find that LLM-based tools, like AutoDroid and ReBL, are effective in inferring the missing operations based on the target app's GUI pages,
but still fail to recognize the redundant operations in the source test.
This leads LLM-based tools to either generate additional actions that mislead the exploration of the target app or get stuck because they fail to find the corresponding actions in the target app.
\par
\textbf{Limitation 2: Ineffective Understanding of Test Case and GUI Pages.}
We find that AutoDroid and ReBL both utilize the DOM attributes of the GUI elements to understand the semantics of the GUI pages.
Also, they use textual representations of the source test to understand the intention of each event.
However, textual representations are often insufficient and ambiguous, leading to difficulties in interpreting the semantics of the source test and the target app's GUI pages.
For example, in Figure~\ref{fig:motivation_example}, the source test script contains a sequence of tap actions to set the bill amount's value,
but based solely on the test script, LLM would infer that the intention of these actions is to set the bill amount to \textbf{5660}, which is incorrect.
\par
\textbf{Limitation 3: Ineffective Termination Mechanism.}
In GUI testing migration, the operation part of the target test should be terminated when the target functionality is triggered.
In fact, AutoDroid and ReBL sometimes fail to recognize the stop condition and continue to generate new action when the target functionality is already triggered, leading to the failure of the migration.
We analyze the root causes for such cases and find that the tools often fail to understand the discrepancies between the source and target app's final states.
In Figure~\ref{fig:motivation_example}, the source app shows the total amount `65.09' as the final state, while the target app shows the total amount in a different format, `\$ 65.09', which is not recognized by ReBL as the same state.

\par
To conclude, both mapping-based and LLM-based tools exhibit limitations in handling the differences in the operational logic between the source and target apps.
Also, both types of tools face challenges in understanding the semantics of the source test and the target app's GUI pages because of the insufficiency and ambiguity of the textual representations they rely on.
Existing LLM-based tools also show limitations in recognizing the stop condition of the target test, leading to non-converging exploration paths.
Motivated by the analysis, we believe that an effective migration approach needs to address the following four aspects:
1) make the mapping in the level of logic steps instead of atomic actions, 
2) eliminate the redundant operations in the source,
3) leverage complementary information in the app to better model and infer the semantics of source test and target app's GUI pages,
4) effectively recognize the stop condition of the target test.

\section{Approach}\label{sec:approach}
\subsection{Overview}
\begin{figure*}[t]
    \centering
    \includegraphics[width=1.0\textwidth]{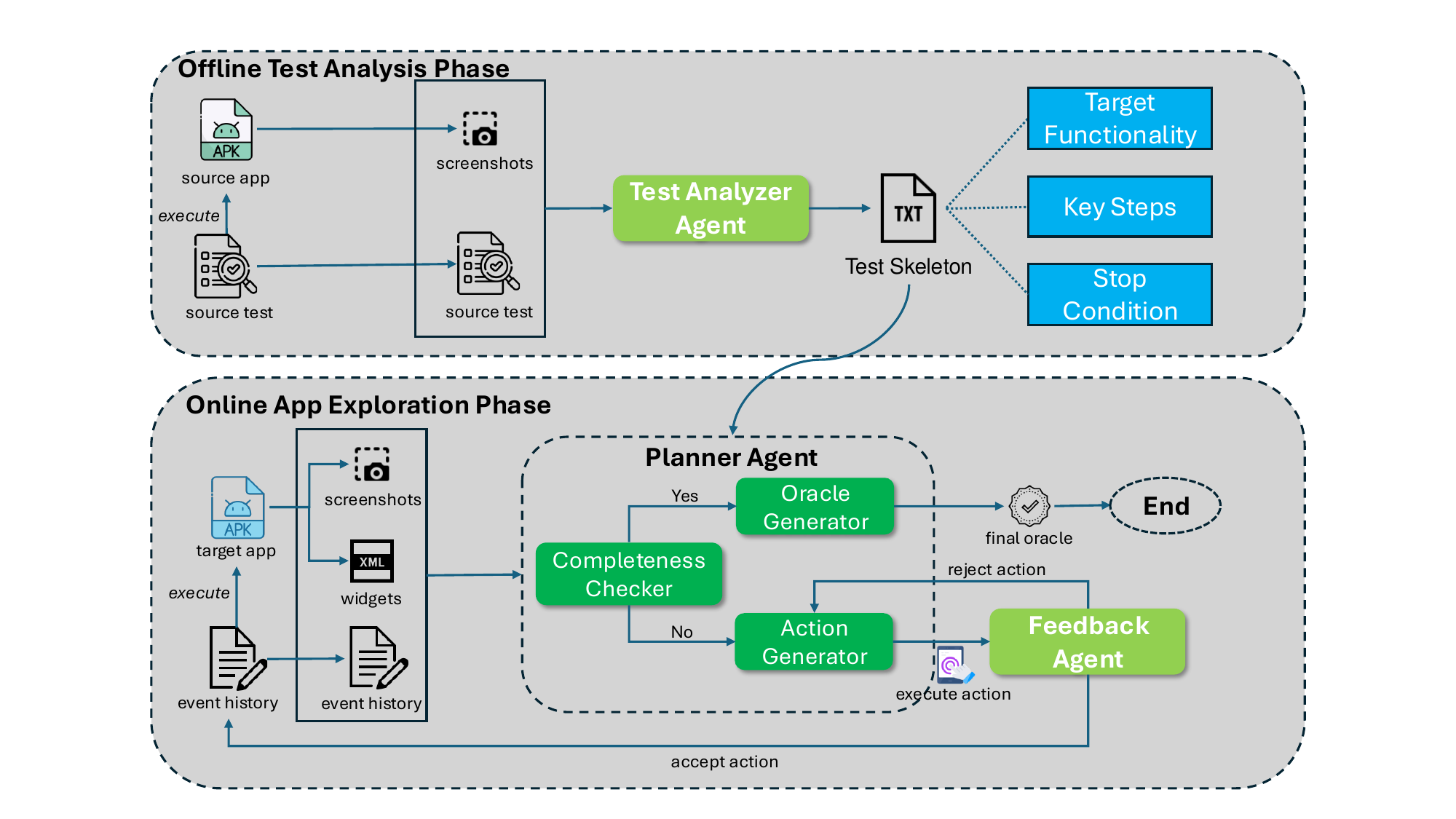}
    \caption{The architecture of \tool}\label{fig:overview}
\end{figure*}
This paper presents \tool, a multi-agent approach for migrating Android GUI tests across apps based on Large Vision-Language Models (VLMs).
\tool employs an exploration-based framework to explore the target app according to the guidance extracted from the source test, aiming at triggering the target functionality on the target app.
The architecture of \tool, as shown in Figure~\ref{fig:overview}, is supported by four agents: \textbf{\taa}, \textbf{\pa}, \textbf{\ea}, and \textbf{\fa}.
The workflow of \tool follows a two-phase process: \textbf{offline test analysis phase} and \textbf{online app exploration phase}.
The offline phase augments the source test with visual execution logs and summarizes a \textbf{test skeleton} from it.
The test skeleton includes the description of the target functionality, the key steps to trigger the functionality, and a stop condition that indicates the successful triggering of the functionality.
Since it has the redundant steps from the source test eliminated, \taa helps prevent misleading exploration in the target app, addressing Limitation 1 of existing LLM-based tools.
The online phase iteratively explores the target app to trigger the target functionality on it, utilizing the test skeleton as soft guidance.
\tool employs \pa to check current exploration progress and generate the next action to take.
In each iteration, \pa dynamically decides the progress of the exploration based on the current GUI state, the event history, and the test skeleton.
\pa also re-evaluates the necessity of each key step in the test skeleton in the dynamic exploration process so that the redundant steps \taa failed to recognize can be further pruned based on the real-time exploration results.
The re-evaluation mechanism further addresses Limitation 1 of existing LLM-based tools.
If the stop condition in the test skeleton is met, and all necessary key steps are completed, \pa generates a closing oracle event to terminate the exploration.
Otherwise, it generates the next action to take based on the current page and the test skeleton.
The mechanism to explicitly evaluate the completeness of key steps effectively addresses Limitation 3 of existing LLM-based tools.
When a new action is generated, \ea executes the action on the target app, captures the execution dialog, and feeds it into \fa.
\fa evaluates the action based on the GUI state changes and the execution dialog and decides whether the action can progress the exploration toward the target functionality.
If so, the action is accepted and added to the target test, and the exploration enters the next iteration.
Otherwise, \fa rejects it and commands \pa to generate another action until an action is accepted.
By integrating visual contexts like app screenshots into \taa, \pa, and \fa, 
\tool utilizes the visual information to provide more comprehensive contexts for the VLMs to understand both the source test and the target app, 
effectively addressing Limitation 2 of existing LLM-based tools.

\subsection{Visual Context Collection}
Three main components of \tool, \taa, \pa, and \fa, are based on VLMs.
As mentioned in Section~\ref{sec:motivation}, the textual attributes of GUI elements\cite{rr13prompting,rr11feedback,onesent2025} and OCR results\cite{temdroid2024} 
used by existing works are often insufficient and ambiguous in expressing the actual semantics of the GUI elements,
and such insufficiency and ambiguity may lead to misinterpretation of the source test and the target app's GUI state.
We assert that visual information serves as an essential complement to textual attributes, given that GUI elements are primarily designed for visual interpretation and human users predominantly comprehend these elements based on their visual appearance.
Therefore, we collect visual information in the form of screenshots of GUI pages to provide more comprehensive contexts to VLMs.
\par
For offline test analysis, we collect screenshots of GUI pages during the source test execution to provide visual materials for the VLMs to understand the intention of each source action.
For online app exploration, the screenshot of the target app's current GUI page is fed into \pa in each iteration so that the agent can understand the current GUI state and generate the next action accordingly.
The screenshots are also utilized by \fa, we provide the screenshots before and after the execution of current action to \fa, so that it can infer the effect of the action based on the GUI state changes.
To summarize, we provide screenshots of GUI pages to VLM-based agents in both offline and online phases as visual contexts to help them understand the source test, the target app, and the execution dialogs.
These screenshots are referred to as visual contexts in the following sections. The screenshots from the source app are called source visual contexts, and the screenshots from the target app are called target visual contexts.

\subsection{\taa}
To address Limitation 1 and Limitation 2 of existing LLM-based tools, we propose \taa to augment the source test with visual execution logs and extract a test skeleton with redundant steps eliminated.
In the workflow, \taa first infers the intention of each source action based on the textual attributes of the source widget and the screenshots taken during the source test execution.
Then, it organizes the atomic actions from the source test into logic steps, 
we do this because sometimes several atomic operations complete a relatively high-level intention, which is more meaningful to the completion of the target functionality,
and more understandable to the VLMs\cite{macdroid2024}.
For example, in Figure~\ref{fig:motivation_example} source test, the atomic operation tapping ``5'' seems meaningless, but when grouping the four tapping operations into a step, 
the intention to input the bill amount is clearer.
After grouping the operations into steps, \taa further categorizes each step into key steps and supporting steps based on their relevance to the target functionality and retains only the key steps.
The key insight here is that the essential logic underlying the target functionality is universally shared across applications, 
while the supporting logic handling app-specific behaviors, may vary across applications, thus, the supporting logic in the source test can be misleading to the exploration in the target app.
For example, in Figure \ref{fig:motivation_example}, the essential logic of the tip calculating functionality is to provide the necessary data, such as the bill amount, for the app to calculate the tip,
while the supporting logic is to handle the tutorial page, confirm input by clicking a button, and anything that is not directly related to the tip calculating functionality but is designed for other purposes, such as enhancing user experience.
We believe by preserving only the key steps, the essential logic embedded in these steps is preserved, while the supporting logic embedded in the supporting steps is eliminated.
Therefore, \taa prevents redundant steps from misleading the exploration process, addressing Limitation 1 of existing LLM-based tools.
\taa also summarizes the target functionality of the source test and determines a stop condition that indicates the end of the operational logic.
All in all, the target functionality, key steps, and stop condition constitute a test skeleton, serving as a soft guidance in the exploration phase.

\subsubsection{Source Action Augmentation}
At the first stage, we utilize a VLM to interpret each source action and its effect on the GUI page into a natural language description, in the format of ``\textit{action} the \textit{widget} to \textit{effect}''.
Specifically, provide VLM with the source test script and its visual execution dialogs in the prompt.
The source test contain the textual attributes of the source widgets and the interaction sequences, VLM can utilize them to understand the intention of each source action.
As shown in Section~\ref{sec: motivation}, the textual attributes are not always sufficient to express the actual semantics of the source widgets, so we add the visual execution dialogs to provide more comprehensive contexts for the VLMs.
The visual execution dialogs contain the screenshots before and after each source action so that the VLMs can understand the actual intention and consequence of each action.
This stage is important for two reasons: 
1) The original source test contains plenty of empty and meaningless textual attributes, 
they contain no useful information but account for a large proportion of the input, 
summarizing the source actions into natural language descriptions can reduce the noise in the input and make the VLMs focus on the essential information\cite{noiserag2024};
2) By incorporating the visual execution dialogs, the ambiguity introduced by insufficient textual attributes can be alleviated in the output of the VLM.
\taa also summarizes the target functionality of the source test, and based on the final oracle event, it determines a stop condition that indicates the end of the operational logic.

\subsubsection{Logic Step Generation}
The second stage is to group the atomic actions into high-level logic steps.
The insight here is that although the source app and target app may activate similar business logic to trigger the target functionality, 
the specific operations to complete the business logic can vary significantly due to the discrepancies in app implementations.
For example, in Figure~\ref{fig:motivation_example}, the source app uses a series of tapping operations to set the bill amount, while the target app may use a text input field.
It is impossible to map each tap operation to the target app directly, but the logic of inputting the bill amount is the same across apps.
Moreover, the logic steps are semantically more meaningful to the VLMs and can help the VLMs understand the intention of the source test better\cite{macdroid2024}.
Therefore, by grouping atomic actions into high-level steps, \taa transforms the output of Stage 1, a sequence of atomic actions,
into a sequence of high-level logic steps, where each step describes a complete intention that relates to the target functionality or app-specific behavior handling.

\subsubsection{Key Step Extraction}
After generating the high-level steps, the \taa further categorizes each step into key steps and supporting steps based on their relevance to the target functionality.
We define key steps as the steps that are directly related to the target functionality itself, and supporting steps as the steps that are used to set up the test environment or handle app-specific behaviors.
Our insight is that the key steps are the essence of the target functionality, and they are shared across the source and target apps, so they cannot be misleading to the exploration process.
Although the supporting steps are app-specific, the supporting steps in the source test can be redundant for the target app and may mislead the exploration process.
We provide the definition of key steps and supporting steps to the VLMs in the prompt, and ask the VLMs to categorize each step into key steps and supporting steps.
As complementary information, we also provide the screenshots before and after each step so that the VLMs can understand the actual effect of each step and make a more accurate decision.
To make sure no key steps are mistakenly removed, \taa queries the VLMs twice and only removes the steps that are identified as supporting steps in both queries.
\par
In short, the \taa summarizes these features from its inputs: the target functionality, the relevant semantic steps, and a stop condition that indicates the end of the test.
These features are then organized as a high-level representation of the source test, which is called the test skeleton, and will be used as a soft guidance for the exploration in the target app.
\begin{itemize}
    \item \textbf{Target Functionality}: \taa infer the target functionality of the source test based on the test scripts and the screenshots.
    \item \textbf{Key Steps}: For each semantic step generalized from actions in the source test, \taa determines whether the step is directly related to the target functionality and only keeps the relevant steps.
    \item \textbf{Stop Condition}: \taa summarizes the semantics of the final oracle event, and uses it as a stop condition that indicates the end of the interaction actions.
\end{itemize}

\begin{table*}[tbp]
    \caption{Prompt Template and Components}\label{tab:prompt_pa}
    \centering
    \scalebox{0.8}{
    \begin{tabular}{|l|l|l|l|}
    \hline
     &                          \textbf{Completeness Checker} & \textbf{Action Generator}  & \textbf{\fa} \\
    \hline
    \textbf{Test Skeleton} & Yes & Yes & Yes \\
    \hline
    \textbf{Source app context} & Final GUI page & No & No \\
    \hline
    \textbf{Target app context} & Page screenshot and description. & \makecell[l]{1. Page description. \\ 2. Pruned Dom Tree. \\ 3. Annotated Screenshot.} & \makecell[l]{Screenshots and descriptions of \\ 1. Page before current action. \\ 2. Page after current action. } \\
    \hline
    \textbf{Event history} & Yes & Yes & Yes \\
    \hline
    \textbf{Chain-of-thought} & \makecell[l]{1. Check step completeness. \\ 2. Check stop condition. \\ 3. Infer navigation to anchor page.} 
                              & \makecell[l]{1. Check step completeness. \\ 2. Infer step undergoing. \\ 3. Infer step to start. \\ 4. Infer connection action to next step.}
                              & \makecell[l]{1. Describe current action consequences. \\ 2. Check action relatedness to functionality.} \\
    \hline
    \textbf{Instruction}      &    \makecell[l]{1. Time to generate oracle? \\ 2. Extra actions to reach anchor page? }
                              &    \makecell[l]{1. Select a widget to interact with. \\ 2. Select an action to perform. }
                              &    \makecell[l]{1. Accept the action or not? \\ 2. Suggest alternative actions.} \\
    \hline
    \end{tabular}
    }
\end{table*}

\subsection{\pa}
The \pa is responsible for deciding when to stop the migration process and generating the next action to take on the target app.
It is composed of three modules: completeness checking (CC), action generation (AG), and oracle generation (OG).
When a migration process starts, in each iteration, \tool dynamically collects textual and visual contexts from the target app, together with the test skeleton, and they are fed into the \pa as inputs.
Once receiving these inputs, the \pa first activates its completeness checking module to determine whether all necessary interactions have been executed to trigger the target app's interested functionality.
If so, the \pa determines that the interaction phase of the target test is complete and then activates its oracle generator to generate a closing oracle event before terminating the migration process.
Otherwise, the \pa activates its action generation module to generate the next action to take on the target app.
\par

\par
\textbf{Completeness Checker}. Completeness Checker determines whether all necessary interactions have been executed to trigger the target app's interested functionality.
CC module takes both textual and visual information of the target app as inputs, including the app's category, the GUI page's real-time screenshot, and the action history.
We also provide the final screen of the source test as a reference so that the CC module can compare the target app's final state with the source test's final state, 
and decide whether the interested functionality has been triggered.
It also integrates the test skeleton and the source app's final state as inputs.
We employ a self-reasoning mechanism for VLMs to think step by step before making a decision. The self-reasoning mechanism proves to be an effective form of chain-of-thought reasoning for VLMs\cite{cot2024}.
Since the core steps are directly related to the interested functionality, we believe all core steps should be executed to trigger the functionality.
The CC module first checks the completeness of each core step in the test skeleton. If all core steps have been completed, the CC module checks whether the stop condition is met by comparing the source app's final state with the target app's real-time state.
If not, it means some extra postfix interactions are needed in the target app to handle app-specific behaviors before triggering the interested functionality.
If any core step is not executed, the CC module double-checks the necessity of this step in the target app by comparing the source app's final state with the target app's real-time state.
Combining the completeness of core steps and stop condition, the CC module determines whether the interaction phase of the target test is complete.

\par
\textbf{Action Generator}. If the CC module determines that the interaction phase is not complete, it activates the AG module to generate the next action to take.
The AG module, based on VLMs, generates the next action by comparing the test skeleton with the target app's textual and visual information.
Like the CC module, the AG module also employs a self-reasoning mechanism for VLMs to generate the next action.
\par
\textbf{Oracle Generator}. If the CC module determines that the interaction phase of the target test is complete, it invokes the OG module to generate a final oracle event and terminate the migration process.
The OG module employs a joint strategy based on both hard rules and VLMs to generate the final oracle event.
When the interaction phase is complete, the OG module first searches the target app's GUI pages for any GUI elements that are identical to the subject GUI element of the final oracle event in the source test.
For example, if the final oracle event in the source test is to check whether a text view displays a specific text, the OG module will search for any text views in the target app that display the same text.
If such a GUI element is found, the OG module generates an oracle event that checks the existence of the GUI element.
Otherwise, the OG module employs VLMs to generate a new oracle event based on the target app's textual and visual information.

\subsection{Multi-Granularity \fa}
\fa is used to provide a self-correction mechanism. 
Due to VLMs' hallucinations and ambiguity in the GUI pages, \pa could make incorrect decisions, and the \fa is responsible for identifying and correcting these errors.
Our \fa operates at two levels of granularity: action level and test level.
\par
At the action level, every time the \pa generates an action, the \fa collects the execution dialog during its execution and assesses the generated action.
At this stage, the \fa assesses the action based on two criteria: whether it is executable and whether it has advanced the test toward triggering the targeted functionality.
If the action is successfully executed, the \fa compares the GUI state before and after the action, analyzing its actual impact.
Based on this analysis, the \fa can decide whether the action has taken effect as expected and progressed the test towards triggering the interested functionality.
If the action has contributed to the test's progress, the \fa accepts it; if not, or if the action fails to execute successfully, the \fa rejects it.
When rejecting an action, the \fa provides a reason for the rejection and prompts the \pa to generate a new action.
This feedback loop continues until the \fa deems the generated action acceptable.
\par
At the test level, the \fa evaluates the entire operational logic of the target test.
Occasionally, the \fa may reject actions consecutively within a single iteration. 
This could occur when an earlier action has directed the exploration to an irrelevant GUI page rather than the current action being faulty.
In such cases, the \fa activates its reflection mechanism to reassess earlier generated actions.
The execution dialog for the currently generated operational logic is collected and fed into the \fa.
The \fa then analyzes the operational logic to identify any earlier actions that may have misled the exploration.
If any misleading actions are found, the \fa truncates the operational logic at the point of the earliest misleading action and prompts the \pa to restart the iteration from that action. 
If no such actions are identified, the \fa will simply instruct the \pa to generate a new action.
\section{Dataset Collection}\label{sec:dataset_collection}
In this section, we introduce the dataset used in this work, \dataset, and the procedure of constructing it.
Specifically, we extend a popular dataset \lin to make it more robust and representative. The extended dataset is called \dataset and is used to evaluate baseline tools and \tool in the following sections.
\par
Several datasets are available for GUI test migration, including those by \lin\cite{craftdroid2019}, Fruiter\cite{fruiter2020}, and ATM\cite{atm2019}, 
all of which are carefully curated and of substantial scale. 
However, a common challenge with these datasets is that they are primarily constructed with old and outdated apps, some of which are no longer installable or maintained\cite{trasm2022}, 
largely reducing the usable portion of the dataset compared to its original size, and thus harming the statistical significance of the evaluation results.
Additionally, the old apps limit the representativeness of the dataset to the current app landscape\cite{appevolution2020}, considering the rapid evolution of Android apps in terms of design philosophy and GUI complexity,
the current datasets may not fully reflect the challenges of GUI test migration on modern apps.
Therefore, we believe it is essential to incorporate newer, more popular apps to create a larger and more representative dataset for evaluating test migration tools.
\begin{figure}[t]
    \centering
    \includegraphics[width=0.5\textwidth]{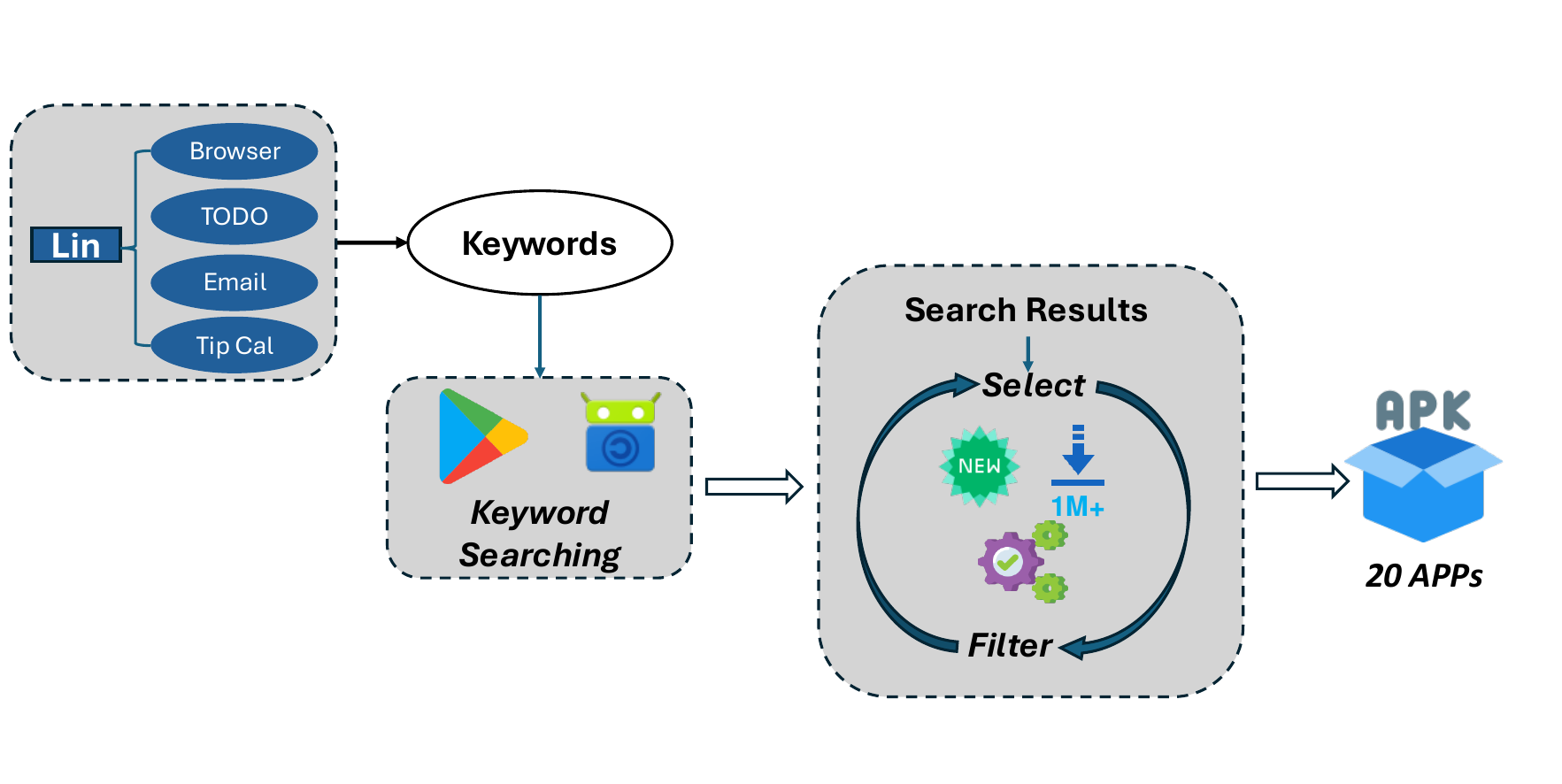}
    \caption{The procedure of dataset collection.}\label{fig:dataset_collection}
\end{figure}
The \lin dataset\cite{craftdroid2019}, one of the most widely used for GUI test migration, consists of 25 apps from 5 categories.
Each category consists of 5 similar apps, and all 5 apps share 2 interested functionalities; with regard to the functionalities, the authors provided two test cases for each app.
Since the tests can be migrated from one app to another as long as they share the same functionalities, the dataset constructs 200 test case pairs for migration in total.
Given the launch dates of the included applications, \lin is partially outdated, with all apps having been released prior to 2019.
Upon review, we found that one application is not installable, four apps fail to function as expected, and tens of test cases are no longer executable.
In this work, we maintain the original structure of the \lin dataset and extend it by adding popular, up-to-date apps to each category.
Also, we update/delete the bad-functioning apps and rewrite non-executable test cases to ensure the robustness and usability of the dataset.
Specifically, we select new apps to boost the dataset, following the criteria:
\begin{itemize}
    \item The app should have been released or updated since 2023/09.
    \item The app must have the functionalities specified by the test cases in \lin dataset.
    \item The app must be either popular on Google Play\cite{googleplay} or open-source on F-Droid\cite{fdroid}.
\end{itemize}
Specifically, we use the category names in \lin as keywords to search for apps on both Google Play and F-Droid. For each category, we collect 3 popular apps from Google Play and 2 open-source apps from F-Droid. 
On Google Play, we select the 3 most downloaded apps from the top 10 search results to ensure their popularity. 
On F-Droid, since download counts are unavailable, we randomly select 2 apps from the top 10 search results. 
If a selected app does not meet the criteria, we exclude it from the search result and perform the selection again until an app that meets the criteria is found.
If we are unable to find enough apps on F-Droid, we supplement the selection with additional apps from Google Play. 

Following this workflow, we collect 15 apps from Google Play and 5 apps from F-Droid, and manually write test cases for each app. The test cases are written in the same format as the \lin dataset.
To ensure precision and rigor, we invite a volunteer with over 2 years of experience in software testing to cross-validate the test cases so that each test case is executable and targets the correct interested functionality.
Additionally, we rewrite the test cases that are not executable. For those that cannot be repaired, we simply delete them from the dataset.
Finally, we get a new dataset consisting of 39 apps across 4 categories and 67 test cases, forming \numMig test case pairs for migration in total. We call this dataset \dataset.
The selected apps are shown in Table~\ref{tab:subject_apps}.
\begin{table}[tbp]
\caption{Subject Apps}\label{tab:subject_apps}
\centering
\resizebox{\columnwidth}{!}{
\begin{tabular}{|l|l|l|}
\hline
\textbf{Category}     & \textbf{App (version)}       & \textbf{Source}   \\ 
\hline
\multirow{10}{*}{a1-Browser}     & a11-Lightning (5.1.0)          & F-Droid        \\
                              & a12-Browser for Android (6.0)  & Google Play    \\
                              & a13-Privacy Browser (2.10)     & F-Droid        \\
                              & a14-FOSS Browser (5.8)         & F-Droid        \\
                              & a15-Firefox Focus (6.0)        & Google Play    \\ 
                              & a16-Chrome (113.0.5672.119)     & Google Play    \\
                              & a17-Opera Mini (80.0)     & Google Play    \\     
                              & a18-Firefox (128.0.2)     & Google Play    \\
                                & a19-DuckDuckGo (5.214.1)    & F-Droid    \\
                                & a10-Fulguris (1.9.30)   & F-Droid    \\ 
\hline

\multirow{10}{*}{a2-To Do List} & a21-Minimal (1.2)              & F-Droid        \\
                              & a22-Clear List (1.5.6)         & F-Droid        \\
                              & a23-To-Do List (2.1)           & F-Droid        \\
                              & a24-Simply Do (0.9.1)          & F-Droid        \\
                              & a25-Shopping List (0.10.1)     & F-Droid        \\
                                & a26-Keep Notes (5.0.411.09.40)    & Google Play    \\
                                & a27-Color Note (4.5.0)    & Google Play    \\
                                & a28-Microsoft To Do (2.114.690.02)    & Google Play    \\
                                & a29-Super Productivity (21.0.0)    & F-Droid    \\
                                & a20-NoNonsense Notes (7.1.7)   & F-Droid    \\ 
\hline

\multirow{9}{*}{a4-Mail Client} & a41-K-9 (5.403)               & Google Play    \\
                              & a42-Email mail box fast mail (1.12.20) & Google Play \\
                              & a43-Mail.Ru (14.120.0)            & Google Play    \\
                              & a44-myMail (7.5.0)             & Google Play    \\
                              & a46-Gmail(2023.04.16.527073575)   & Google Play    \\
                              & a47-Yahoo Mail (7.39.0)   & Google Play    \\
                              & a48-Microsoft Outlook (4.2415.1)  & Google Play    \\
                              & a49-BlueMail (1.9.49)  & Google Play    \\
                              & a40-Monocles Mail (1.2.3)  & F-Droid    \\
\hline
\multirow{10}{*}{a5-Tip Calculator} & a51-Tip Calculator (1.1)    & Google Play    \\
                              & a52-Tip Calc (1.11)            & Google Play    \\
                              & a53-Simple Tip Calculator (1.2) & Google Play   \\
                              & a54-Tip Calculator Plus (2.0)  & Google Play    \\
                              & a55-Free Tip Calculator (1.0.0.9) & Google Play \\
                              & a56-Tip N Split (2.0.6) & Google Play    \\
                              & a57-Flatiron Tip Calculator (3.1.2) & Google Play    \\
                              & a58-Bitskon Tip Calculator (5.0) & Google Play    \\
                              & a59-Chimbori Tip Calculator (7.5.0) & Google Play    \\
                              & a50-TuriApps Tip Calculator (4.0.0) & Google Play    \\
\hline

\end{tabular}
}
\end{table}

\section{Evaluation}\label{sec:evaluation}
To comprehensively evaluate \tool's effectiveness and propose suggestions for future research, we conduct extensive experiments addressing four key aspects:
The effectiveness of \tool in GUI test migration (RQ1), the contribution of main components on effectiveness (RQ2), 
and the reasons for failed migrations (RQ3).

\subsection{Experimental Setup}
\subsubsection{Study Design}
We conducted experiments to answer the following research questions:
\begin{itemize}
    \item \textbf{RQ1.} What are the effectiveness and efficiency of \tool compared to existing approaches?
    \item \textbf{RQ2.} What is the contribution of the main components of \tool on the migration performance?
    % \item \textbf{RQ3.} How useful is \tool on out-of-domain apps in real-world scenarios?
    \item \textbf{RQ3.} What are the reasons for the failed migrations performed by \tool?
\end{itemize}
\subsubsection{Baselines}
We use the three subject tools introduced in Section~\ref{sec: motivation} as baselines for comparison because they are state-of-the-art and representative.
Since AutoDroid is for UI task automation and ReBL is for bug reproduction,
we adapt them to the test migration task by providing the source test's natural language descriptions as input. The descriptions are summarized by GPT-4o and reviewed by a volunteer with over 2 years of experience in software testing.
The natural language descriptions are regarded by AutoDroid as user instructions and by ReBL as bug reports to reproduce.
To ensure experimental fairness and consistency, we use GPT-4o to power both LLM-based subject tools in this study since it is one of the most popular and powerful LLMs available.
\subsubsection{Metrics}
We use the success rate of the migration as the evaluation metric for this study.
Specifically, if a generated test case is executable and examines the correct target functionality of the target app, it is regarded as successful,
and the success rate is the proportion of successful test cases among all generated test cases.
\par
To get the success rate while involving human intervention as little as possible,
we set up a 4-step verification rule to determine whether a test case examines the correct functionality, and ensure that any test case surviving the whole verification process is successful, and any test case failing in any step is labeled as failed.
We first filter out the test cases that are not executable and label them as failed.
Then, we compare the generated test cases with the ground truth test cases. If a generated test case is identical to a ground truth test case, it is labeled successful.
Thirdly, for executable test cases that are not identical to the ground truth, we activate a manually written evaluator that checks the final GUI state of the target app after the test execution.
Specifically, the evaluator looks up the page layouts of the target app recorded during the test execution and seeks the anchor widget specified in the grout truth test case's oracle event.
If the anchor widget is found, it means the test case leads to the correct GUI state, thus triggering the correct functionality.
Fourthly, we manually examine the ones that are not verified by the evaluator. We check the execution logs to determine whether the test case examines the correct functionality,
The manual examination step is necessary for eliminating false negatives because the evaluator may fail to locate the anchor widget due to the complexity of the GUI and dynamic features of the app.
\subsubsection{Dataset}
We use the \dataset dataset introduced in Section~\ref{sec:dataset_collection} for both RQ1 and RQ2.

\subsubsection{Hardware}
All experiments are performed on two Android emulators, Pixel 5 API 23 and Pixel 5 API 34, both running on a 16-inch MacBook Pro (2021) equipped with an M1 Pro chip and 16GB of RAM.
By default, the apps from the original \lin dataset run on the API 23 emulator, and the apps newly added in \dataset run on the API 34 emulator.

\begin{table*}[tbp]
    \caption{Success Rate of Baselines and ReuseDroid on \dataset}\label{tab:overall_result}
    \centering
    \begin{tabular}{lrrrrr}
    \toprule
    % \textbf{Approach} & \multicolumn{1}{c}{\textbf{Browser}} & \multicolumn{1}{c}{\textbf{ToDo List}} & \multicolumn{1}{c}{\textbf{Mail Client}} & \multicolumn{1}{c}{\textbf{Tip Calculator}} & \multicolumn{1}{c}{\textbf{All Categories}} \\
    \textbf{Approach} & \textbf{Browser} & \textbf{ToDo List} & \textbf{Mail Client} & \textbf{Tip Calculator} & \textbf{All Categories} \\
    \midrule
    \textbf{CraftDroid}  & 33.3\%             & 14.4\%               & 35.7\%                 & 15.0\%                   & 21.6\%                     \\
    \midrule
    \textbf{ReBL-GPT-4o}  & 16.2\%             & 49.7\%               & 34.4\%                 & 71.2\%                   & 43.1\%                     \\
    \midrule
    \textbf{Autodroid-GPT-4o} & 17.0\%          & 38.3\%               & 12.5\%                 & 32.2\%                   & 22.2\%                     \\
    \midrule
    \textbf{ReuseDroid-GPT-4o}        & \textcolor{red}{95.2\%}         & \textcolor{red}{80.9\%} & 97.6\% & \textcolor{red}{92.8\%} & \textcolor{red}{\textbf{90.3\%}} \\
    \midrule
    \textbf{ReuseDroid-qwen-vl-max}   & 90.1\%    & 76.3\% & \textcolor{red}{98.2\%} & 87.7\% & 86.5\% \\
    \bottomrule
    \end{tabular}
\end{table*}

\subsection{Results and Analysis}
\subsubsection{RQ1. What is the effectiveness and efficiency of \tool compared to existing approaches?}
The results of RQ1 are shown in Table~\ref{tab:overall_result}.
\tool powered by GPT-4o achieves the highest overall success rate of 90.3\%, surpassing the best-performing baseline ReBL by 109.5\%.
\tool powered by the open-source model qwen-vl-max also outperforms all the baselines, with a success rate of 86.5\%.
It is observed that \tool performs significantly poorer on the category \textit{ToDo List}. We analyze the failed cases and find that the main reason is the presence of ambiguous UI elements and complex operations in the apps, which confuse VLMs to make correct decisions.
The results demonstrate the effectiveness of \tool in GUI test migration.

We also compare the efficiency of \tool and the baselines.
On average, \tool takes 279.9 seconds to complete a migration.
Therefore, our multi-agent workflow is comparable to AutoDroid's 276.3 seconds and ReBL's 243.7 seconds in terms of efficiency.
All the LLM-based tools are much more efficient than CraftDroid, which takes 1.5 hours on average to complete a migration.

\begin{table}[tbp]
\caption{Contribution of different Components}\label{tab:ablation_study}
\centering
\begin{tabular}{lr}
\toprule
\textbf{Approach} & \textbf{Success Rate} \\
\midrule
\textbf{ReuseDroid-GPT-4o}               & 90.3\% \\
\textbf{ReuseDroid-GPT-4o w/o Vision}    & 78.2\%  \\
\textbf{ReuseDroid-GPT-4o w/o TAA}        & 74.6\%  \\
\textbf{ReuseDroid-GPT-4o w/o Feedback}  & 82.5\%  \\
\bottomrule
\end{tabular}
\end{table}

\subsubsection{RQ2. What is the impact of different components of \tool on the migration performance?}
e include three components of \tool in the ablation study and evaluate the contribution of each component to the migration performance.
Specifically, we investigate the impact of visual contexts, \taa, and \fa, on the migration success rate.
The three components are disabled one by one so that the impact of each component can be reflected by the change in the migration performance.
\tool with each component disabled is denoted as \textit{\tool w/o Vision}, \textit{\tool w/o TAA}, and \textit{\tool w/o Feedback}, respectively.
\par
The results of the ablation study are shown in Table~\ref{tab:ablation_study}.
We observe that all three components largely contribute to the migration performance.
Among the variants, \tool w/o \taa performs the worst, with a success rate of 74.6\%.
This result indicates that \taa is essential for \tool to generate accurate and complete test cases, 
considering it not only augments the source test with visual contexts but also generalizes atomic actions to logic steps and eliminates redundant ones.
The results also suggest that the visual contexts and feedback mechanism are important for \tool to understand the GUI pages and to correct the generated test cases, respectively.
Without the visual contexts, \tool can be confused by the meaningless, empty, and ambiguous textual attributes in the source test and DOM tree.
Without the feedback mechanism, \tool generates actions based on the GUI state before the action execution, without observing its actual effect on the GUI state.

\subsubsection{RQ3. What are the reasons for the failed migrations performed by \tool?}
\begin{figure}
    \centering
    \includegraphics[width=0.3\textwidth]{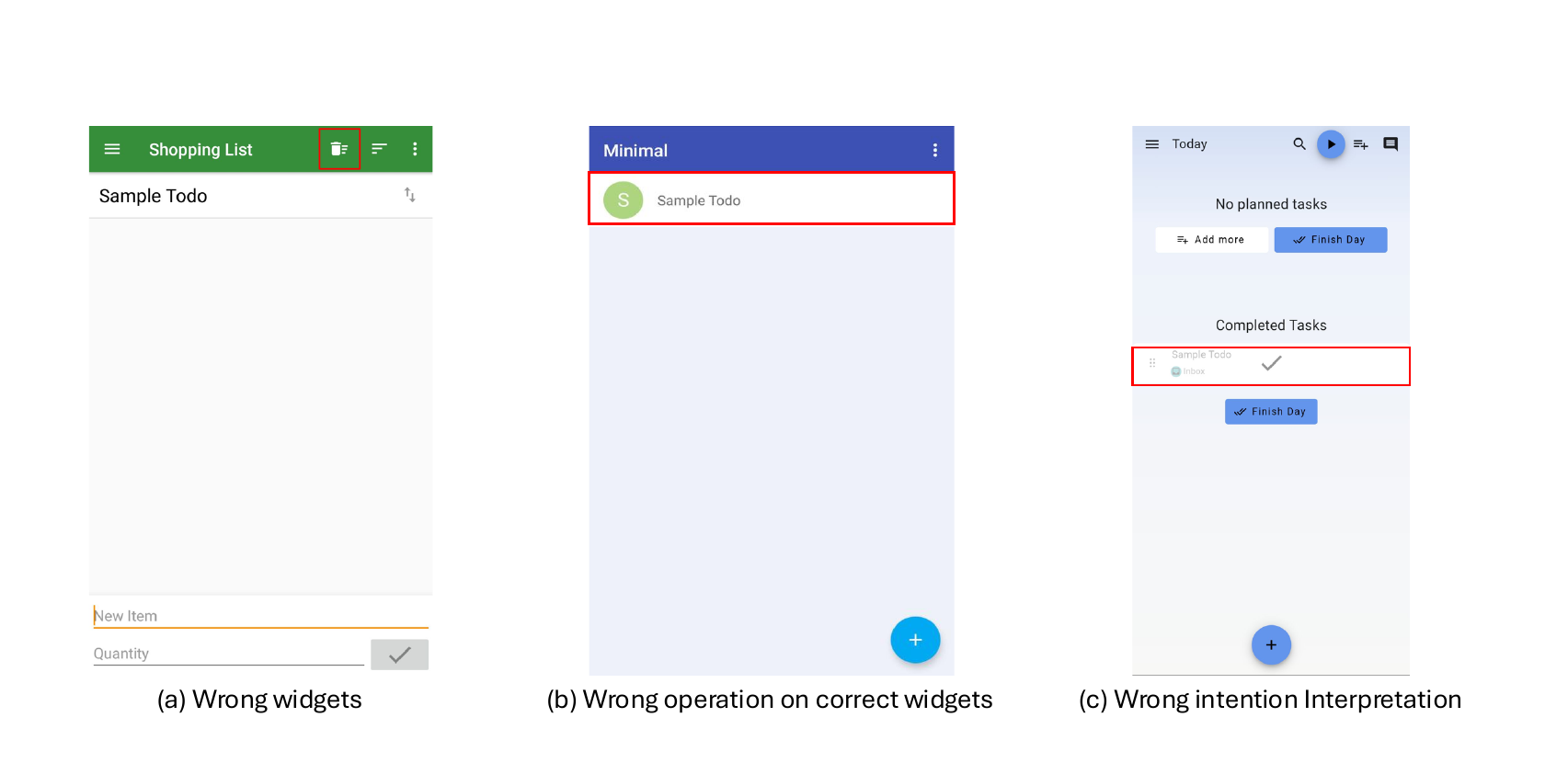}
    \caption{Failed cases for deleting ``Sample Todo''}\label{fig:failed_case_example}
\end{figure}
To systematically analyze the limitations of \tool, we examine the failed migration cases and identify three primary categories of failures.
We illustrate these challenges using a representative example of deleting a ``Sample Todo'' in the \textit{ToDo List} category.
\par
\textbf{Ambiguous UI Elements.} 
Despite the incorporation of visual contexts to enhance GUI page understanding,
certain widgets exhibit inherent ambiguity in both their textual attributes and visual representations.
As demonstrated in Figure~\ref{fig:failed_case_example}(a), the `delete' button is specifically designed to clear all completed todos rather than targeting individual items like the ``Sample Todo''.
The button's visual appearance and placement, however, mislead \tool into incorrectly identifying it as the appropriate target for single-item deletion.
\par
\textbf{Non-intuitive Operations.}
The interaction patterns with widgets can be complex and non-intuitive.
In Figure~\ref{fig:failed_case_example}(b), the correct operation to delete the ``Sample Todo'' is to swipe it to the left,
but \tool incorrectly attempts to long-click it, likely due to the VLMs' preconceived notion that items are typically deleted through long-press actions.
\par
\textbf{Misinterpretation of Target Functionality.}
\tool occasionally misinterprets the intended functionality of the source test.
The action depicted in Figure~\ref{fig:failed_case_example}(c) fails
because \tool incorrectly identifies the target functionality as `mark as completed' when the actual goal is to `delete'.
\section{Threats to Validity}\label{sec:threats_to_validity}
Our study is subject to several threats to validity.
The internal validity threats are related to the correctness of manually written test cases during dataset construction and the correctness of successful test case labeling during evaluation.
To mitigate the threats introduced by manually written test cases, we invite an experienced undergraduate student with more than 2 years of experience in software testing to cross-validate the test cases.
We also employ a 4-step verification process to label the generated test cases and invite three students experienced in Android app testing to cross-verify the test cases. In doing so, the unreliability introduced by manual verification is minimized.

The external validity threats are related to the randomness and selection of VLMs empowering \tool and baselines, 
we mitigate the former by running \tool and baselines three times and reporting the average performance.
For VLM selection, we choose two models, GPT-4o and GPT-4o-mini, from the most popular GPT family.
To ensure the generalizability of the results, we also include an open-source model, qwen-vl-max.

\section{Related Work}\label{sec:related_work}
\textbf{GUI Test Migration.} There are several approaches for migrating GUI tests across apps, which can be categorized into two types: mapping-based and exploration-based.
Earlier works employ a mapping-based framework\cite{appflow2018,atm2019,craftdroid2019,temdroid2024}, where each source action, according to the semantics of its corresponding GUI element, is assigned to a target widget.
Such works map source widgets to target widgets based on their semantic similarity, which is calculated by embedding the GUI elements into a feature space.
AppFlow\cite{appflow2018} uses a GUI element's DOM attributes, sizes, positions, image icons, and neighboring elements as features to represent its semantics.
These features are then fed into a support vector machine (SVM) to produce embeddings for the GUI elements.
ATM\cite{atm2019} and CraftDroid\cite{craftdroid2019} employ a word2vec model\cite{word2vec} trained on 500 GUI test cases and Google News, respectively, which turns out to be more effective than AppFlow's SVM.
TEMDroid\cite{temdroid2024} trains a BERT model\cite{bert} to encode the source test case and the target app's GUI structure. The authors make the training possible by constructing a high-quality widget-widget mapping dataset from Semfinder\cite{semfinder2021}.
For mapping-based approaches, although the tools are allowed to explore the target app's GUI from time to time, such explorations are aimed at finding the target widgets that match the source widgets rather than understanding the target app's functionalities and finding an appropriate operational logic.
In comparison, MACdroid\cite{macdroid2024} is the first work that uses large language models for GUI test migration. The tool automatically prompts LLM to explore the target app's GUI pages, trigger the target functionality, and generate the interaction sequences accordingly.
The authors also utilize LLM to abstract the source test case instead of using the atomic operations directly, making a high-level representation of the source test case.
\par
\textbf{Bug Report Reproduction.} Bug report reproduction is a task that reproduces bugs in an app based on a bug report collected from user feedback.
By providing a bug report, a user actually provides a high-level description of the basic steps to reproduce the bug, so bug report reproduction is similar to GUI test migration by migrating a user-provided operational logic to the real app and generating an executable test case.
Some existing approaches\cite{rr1auto,rr2rec,rr5rec} adopt a similar mapping-based framework as earlier GUI test migration techniques, in which a tool first identifies step-to-reproduce(S2R) entities in the bug report and then matches each of them to the corresponding GUI elements in the app.
Fazzini et al.\cite{rr1auto} define a set of keywords that indicate the presence of S2Rs in the bug reports and then use keyword searching to identify the sections (usually paragraphs) containing the keywords.
However, their approach only achieves a coarse-grained identification since even in the same section, some sentences may not be related to the S2Rs.
To address this issue, Zhao et al.\cite{rr2rec} manually review 813 bug reports and identify 22 common grammar patterns used for describing the S2Rs, among them 7 are for ``click'' actions, 14 are for ``input'' actions, and 1 is for ``gesture'' actions.
They then use SpaCy to parse sentences in the bug reports into structured dependency trees and compare the dependency trees with the predefined grammar patterns to identify the sentences containing the S2Rs.
These methods rely on predefined grammar patterns to identify the S2Rs, which may not be able to handle the diverse expressions in the bug reports.
Zhao et al.\cite{rr5rec} propose a method where a neural network is trained to label whether a sentence in the bug report is an S2R, eliminating the dependency on predefined grammar patterns.
In more recent works, LLMs have been used to identify the S2Rs in bug reports.
Feng et al.\cite{rr13prompting} define a domain-specific language (DSL) to regulate LLMs' outputs.
ReBL is then proposed by Wang et al.\cite{rebl2024}, where the non-S2R information in the bug report is also used to guide the exploration process.
ReBL adopts an exploration-based framework to reproduce bugs and follows a feedback-driven exploration strategy to guide the LLMs in exploring the application and triggering the specified bug.
\par
\textbf{LLM for UI Automation.} LLMs have been widely used in UI automation tasks, such as generating UI automation scripts from user instructions.
Zhe et al.\cite{gm1expert} prompt the LLM with the current GUI state, the application's critical information like the application's name and current activity's name, and expect the LLM to generate the next action to take.
Yoon et al.\cite{gm2intent} propose a feedback-enhanced strategy to validate the interaction sequences generated by the LLMs and correct potential wrong actions.
\section{Conclusion}
In conclusion, this paper introduces a multi-VLM framework for GUI test migration, 
which systematically analyzes the source test, evaluates the target application, and generates actions and feedback iteratively.
The framework overcomes the limitations of existing approaches in managing divergent operation logic and complex GUI interactions.
Experimental results highlight the effectiveness and efficiency of \tool while identifying opportunities for future enhancements, 
such as addressing non-intuitive operations and ambiguous UI elements.

\balance

\bibliographystyle{IEEEtran}
\bibliography{references}

\begin{thebibliography}{10}
\providecommand{\url}[1]{#1}
\csname url@samestyle\endcsname
\providecommand{\newblock}{\relax}
\providecommand{\bibinfo}[2]{#2}
\providecommand{\BIBentrySTDinterwordspacing}{\spaceskip=0pt\relax}
\providecommand{\BIBentryALTinterwordstretchfactor}{4}
\providecommand{\BIBentryALTinterwordspacing}{\spaceskip=\fontdimen2\font plus
\BIBentryALTinterwordstretchfactor\fontdimen3\font minus \fontdimen4\font\relax}
\providecommand{\BIBforeignlanguage}[2]{{%
\expandafter\ifx\csname l@#1\endcsname\relax
\typeout{** WARNING: IEEEtran.bst: No hyphenation pattern has been}%
\typeout{** loaded for the language `#1'. Using the pattern for}%
\typeout{** the default language instead.}%
\else
\language=\csname l@#1\endcsname
\fi
#2}}
\providecommand{\BIBdecl}{\relax}
\BIBdecl

\bibitem{stoat2017}
\BIBentryALTinterwordspacing
T.~Su, G.~Meng, Y.~Chen, K.~Wu, W.~Yang, Y.~Yao, G.~Pu, Y.~Liu, and Z.~Su, ``Guided, stochastic model-based gui testing of android apps,'' in \emph{Proceedings of the 2017 11th Joint Meeting on Foundations of Software Engineering}, ser. ESEC/FSE 2017.\hskip 1em plus 0.5em minus 0.4em\relax New York, NY, USA: Association for Computing Machinery, 2017, p. 245–256. [Online]. Available: \url{https://doi.org/10.1145/3106237.3106298}
\BIBentrySTDinterwordspacing

\bibitem{atm2019}
F.~Behrang and A.~Orso, ``Test migration between mobile apps with similar functionality,'' in \emph{2019 34th IEEE/ACM International Conference on Automated Software Engineering (ASE)}, 2019, pp. 54--65.

\bibitem{craftdroid2019}
\BIBentryALTinterwordspacing
J.-W. Lin, R.~Jabbarvand, and S.~Malek, ``Test transfer across mobile apps through semantic mapping,'' in \emph{Proceedings of the 34th IEEE/ACM International Conference on Automated Software Engineering}, ser. ASE '19.\hskip 1em plus 0.5em minus 0.4em\relax IEEE Press, 2020, p. 42–53. [Online]. Available: \url{https://doi.org/10.1109/ASE.2019.00015}
\BIBentrySTDinterwordspacing

\bibitem{temdroid2024}
\BIBentryALTinterwordspacing
Y.~Zhang, W.~Zhang, D.~Ran, Q.~Zhu, C.~Dou, D.~Hao, T.~Xie, and L.~Zhang, ``Learning-based widget matching for migrating gui test cases,'' in \emph{Proceedings of the IEEE/ACM 46th International Conference on Software Engineering}, ser. ICSE '24.\hskip 1em plus 0.5em minus 0.4em\relax New York, NY, USA: Association for Computing Machinery, 2024. [Online]. Available: \url{https://doi.org/10.1145/3597503.3623322}
\BIBentrySTDinterwordspacing

\bibitem{migratepro2024}
\BIBentryALTinterwordspacing
Y.~Zhang, Q.~Zhu, J.~Yan, C.~Liu, W.~Zhang, Y.~Zhao, D.~Hao, and L.~Zhang, ``Synthesis-based enhancement for gui test case migration,'' in \emph{Proceedings of the 33rd ACM SIGSOFT International Symposium on Software Testing and Analysis}, ser. ISSTA 2024.\hskip 1em plus 0.5em minus 0.4em\relax New York, NY, USA: Association for Computing Machinery, 2024, p. 869–881. [Online]. Available: \url{https://doi.org/10.1145/3650212.3680327}
\BIBentrySTDinterwordspacing

\bibitem{word2vec}
\BIBentryALTinterwordspacing
T.~Mikolov, K.~Chen, G.~Corrado, and J.~Dean, ``Efficient estimation of word representations in vector space,'' 2013. [Online]. Available: \url{https://arxiv.org/abs/1301.3781}
\BIBentrySTDinterwordspacing

\bibitem{bert}
\BIBentryALTinterwordspacing
J.~Devlin, M.-W. Chang, K.~Lee, and K.~Toutanova, ``Bert: Pre-training of deep bidirectional transformers for language understanding,'' 2019. [Online]. Available: \url{https://arxiv.org/abs/1810.04805}
\BIBentrySTDinterwordspacing

\bibitem{sail2024}
\BIBentryALTinterwordspacing
M.~Wu, H.~Wang, J.~Ren, Y.~Cao, Y.~Li, A.~Jiang, D.~Ran, Y.~Hu, W.~Yang, and T.~Xie, ``Skill-adpative imitation learning for ui test reuse,'' 2024. [Online]. Available: \url{https://arxiv.org/abs/2409.13311}
\BIBentrySTDinterwordspacing

\bibitem{rebl2024}
\BIBentryALTinterwordspacing
D.~Wang, Y.~Zhao, S.~Feng, Z.~Zhang, W.~G.~J. Halfond, C.~Chen, X.~Sun, J.~Shi, and T.~Yu, ``Feedback-driven automated whole bug report reproduction for android apps,'' in \emph{Proceedings of the 33rd ACM SIGSOFT International Symposium on Software Testing and Analysis}, ser. ISSTA 2024.\hskip 1em plus 0.5em minus 0.4em\relax New York, NY, USA: Association for Computing Machinery, 2024, p. 1048–1060. [Online]. Available: \url{https://doi.org/10.1145/3650212.3680341}
\BIBentrySTDinterwordspacing

\bibitem{autodroid2024}
\BIBentryALTinterwordspacing
H.~Wen, Y.~Li, G.~Liu, S.~Zhao, T.~Yu, T.~J.-J. Li, S.~Jiang, Y.~Liu, Y.~Zhang, and Y.~Liu, ``Autodroid: Llm-powered task automation in android,'' in \emph{Proceedings of the 30th Annual International Conference on Mobile Computing and Networking}, ser. ACM MobiCom '24.\hskip 1em plus 0.5em minus 0.4em\relax New York, NY, USA: Association for Computing Machinery, 2024, p. 543–557. [Online]. Available: \url{https://doi.org/10.1145/3636534.3649379}
\BIBentrySTDinterwordspacing

\bibitem{rr13prompting}
\BIBentryALTinterwordspacing
S.~Feng and C.~Chen, ``Prompting is all you need: Automated android bug replay with large language models,'' in \emph{Proceedings of the IEEE/ACM 46th International Conference on Software Engineering}, ser. ICSE '24.\hskip 1em plus 0.5em minus 0.4em\relax New York, NY, USA: Association for Computing Machinery, 2024. [Online]. Available: \url{https://doi.org/10.1145/3597503.3608137}
\BIBentrySTDinterwordspacing

\bibitem{rr11feedback}
\BIBentryALTinterwordspacing
D.~Wang, Y.~Zhao, S.~Feng, Z.~Zhang, W.~G.~J. Halfond, C.~Chen, X.~Sun, J.~Shi, and T.~Yu, ``Feedback-driven automated whole bug report reproduction for android apps,'' in \emph{Proceedings of the 33rd ACM SIGSOFT International Symposium on Software Testing and Analysis}, ser. ISSTA 2024.\hskip 1em plus 0.5em minus 0.4em\relax New York, NY, USA: Association for Computing Machinery, 2024, p. 1048–1060. [Online]. Available: \url{https://doi.org/10.1145/3650212.3680341}
\BIBentrySTDinterwordspacing

\bibitem{onesent2025}
Y.~Huang, J.~Wang, Z.~Liu, M.~Li, S.~Wang, C.~Chen, Y.~Hu, and Q.~Wang, ``One sentence can kill the bug: Auto-replay mobile app crashes from one-sentence overviews,'' \emph{IEEE Transactions on Software Engineering}, pp. 1--15, 2025.

\bibitem{macdroid2024}
\BIBentryALTinterwordspacing
Y.~Zhang, C.~Liu, X.~Xie, Y.~Lin, J.~S. Dong, D.~Hao, and L.~Zhang, ``Llm-based abstraction and concretization for gui test migration,'' 2024. [Online]. Available: \url{https://arxiv.org/abs/2409.05028}
\BIBentrySTDinterwordspacing

\bibitem{noiserag2024}
\BIBentryALTinterwordspacing
F.~Cuconasu, G.~Trappolini, F.~Siciliano, S.~Filice, C.~Campagnano, Y.~Maarek, N.~Tonellotto, and F.~Silvestri, ``The power of noise: Redefining retrieval for rag systems,'' in \emph{Proceedings of the 47th International ACM SIGIR Conference on Research and Development in Information Retrieval}, ser. SIGIR '24.\hskip 1em plus 0.5em minus 0.4em\relax New York, NY, USA: Association for Computing Machinery, 2024, p. 719–729. [Online]. Available: \url{https://doi.org/10.1145/3626772.3657834}
\BIBentrySTDinterwordspacing

\bibitem{cot2024}
X.~Zhang, C.~Du, T.~Pang, Q.~Liu, W.~Gao, and M.~Lin, ``Chain of preference optimization: Improving chain-of-thought reasoning in llms,'' in \emph{Advances in Neural Information Processing Systems}, A.~Globerson, L.~Mackey, D.~Belgrave, A.~Fan, U.~Paquet, J.~Tomczak, and C.~Zhang, Eds., vol.~37.\hskip 1em plus 0.5em minus 0.4em\relax Curran Associates, Inc., 2024, pp. 333--356.

\bibitem{fruiter2020}
\BIBentryALTinterwordspacing
Y.~Zhao, J.~Chen, A.~Sejfia, M.~Schmitt~Laser, J.~Zhang, F.~Sarro, M.~Harman, and N.~Medvidovic, ``Fruiter: a framework for evaluating ui test reuse,'' in \emph{Proceedings of the 28th ACM Joint Meeting on European Software Engineering Conference and Symposium on the Foundations of Software Engineering}, ser. ESEC/FSE 2020.\hskip 1em plus 0.5em minus 0.4em\relax New York, NY, USA: Association for Computing Machinery, 2020, p. 1190–1201. [Online]. Available: \url{https://doi.org/10.1145/3368089.3409708}
\BIBentrySTDinterwordspacing

\bibitem{trasm2022}
S.~Liu, Y.~Zhou, T.~Han, and T.~Chen, ``Test reuse based on adaptive semantic matching across android mobile applications,'' in \emph{2022 IEEE 22nd International Conference on Software Quality, Reliability and Security (QRS)}, 2022, pp. 703--709.

\bibitem{appevolution2020}
\BIBentryALTinterwordspacing
T.~Li, M.~Zhang, H.~Cao, Y.~Li, S.~Tarkoma, and P.~Hui, ``”what apps did you use?”: Understanding the long-term evolution of mobile app usage,'' in \emph{Proceedings of The Web Conference 2020}, ser. WWW '20.\hskip 1em plus 0.5em minus 0.4em\relax New York, NY, USA: Association for Computing Machinery, 2020, p. 66–76. [Online]. Available: \url{https://doi.org/10.1145/3366423.3380095}
\BIBentrySTDinterwordspacing

\bibitem{googleplay}
\BIBentryALTinterwordspacing
``Google play store.'' [Online]. Available: \url{https://play.google.com/store}
\BIBentrySTDinterwordspacing

\bibitem{fdroid}
\BIBentryALTinterwordspacing
``F-droid.'' [Online]. Available: \url{https://f-droid.org}
\BIBentrySTDinterwordspacing

\bibitem{appflow2018}
\BIBentryALTinterwordspacing
G.~Hu, L.~Zhu, and J.~Yang, ``Appflow: using machine learning to synthesize robust, reusable ui tests,'' in \emph{Proceedings of the 2018 26th ACM Joint Meeting on European Software Engineering Conference and Symposium on the Foundations of Software Engineering}, ser. ESEC/FSE 2018.\hskip 1em plus 0.5em minus 0.4em\relax New York, NY, USA: Association for Computing Machinery, 2018, p. 269–282. [Online]. Available: \url{https://doi.org/10.1145/3236024.3236055}
\BIBentrySTDinterwordspacing

\bibitem{semfinder2021}
\BIBentryALTinterwordspacing
L.~Mariani, A.~Mohebbi, M.~Pezz\`{e}, and V.~Terragni, ``Semantic matching of gui events for test reuse: are we there yet?'' in \emph{Proceedings of the 30th ACM SIGSOFT International Symposium on Software Testing and Analysis}, ser. ISSTA 2021.\hskip 1em plus 0.5em minus 0.4em\relax New York, NY, USA: Association for Computing Machinery, 2021, p. 177–190. [Online]. Available: \url{https://doi.org/10.1145/3460319.3464827}
\BIBentrySTDinterwordspacing

\bibitem{rr1auto}
\BIBentryALTinterwordspacing
M.~Fazzini, M.~Prammer, M.~d'Amorim, and A.~Orso, ``Automatically translating bug reports into test cases for mobile apps,'' in \emph{Proceedings of the 27th ACM SIGSOFT International Symposium on Software Testing and Analysis}, ser. ISSTA 2018.\hskip 1em plus 0.5em minus 0.4em\relax New York, NY, USA: Association for Computing Machinery, 2018, p. 141–152. [Online]. Available: \url{https://doi.org/10.1145/3213846.3213869}
\BIBentrySTDinterwordspacing

\bibitem{rr2rec}
Y.~Zhao, T.~Yu, T.~Su, Y.~Liu, W.~Zheng, J.~Zhang, and W.~G.J.~Halfond, ``Recdroid: Automatically reproducing android application crashes from bug reports,'' in \emph{2019 IEEE/ACM 41st International Conference on Software Engineering (ICSE)}, 2019, pp. 128--139.

\bibitem{rr5rec}
\BIBentryALTinterwordspacing
Y.~Zhao, T.~Su, Y.~Liu, W.~Zheng, X.~Wu, R.~Kavuluru, W.~G.~J. Halfond, and T.~Yu, ``Recdroid+: Automated end-to-end crash reproduction from bug reports for android apps,'' \emph{ACM Trans. Softw. Eng. Methodol.}, vol.~31, no.~3, Mar. 2022. [Online]. Available: \url{https://doi.org/10.1145/3488244}
\BIBentrySTDinterwordspacing

\bibitem{gm1expert}
\BIBentryALTinterwordspacing
Z.~Liu, C.~Chen, J.~Wang, M.~Chen, B.~Wu, X.~Che, D.~Wang, and Q.~Wang, ``Make llm a testing expert: Bringing human-like interaction to mobile gui testing via functionality-aware decisions,'' in \emph{Proceedings of the IEEE/ACM 46th International Conference on Software Engineering}, ser. ICSE '24.\hskip 1em plus 0.5em minus 0.4em\relax New York, NY, USA: Association for Computing Machinery, 2024. [Online]. Available: \url{https://doi.org/10.1145/3597503.3639180}
\BIBentrySTDinterwordspacing

\bibitem{gm2intent}
\BIBentryALTinterwordspacing
J.~Yoon, R.~Feldt, and S.~Yoo, ``{ Intent-Driven Mobile GUI Testing with Autonomous Large Language Model Agents },'' in \emph{2024 IEEE Conference on Software Testing, Verification and Validation (ICST)}.\hskip 1em plus 0.5em minus 0.4em\relax Los Alamitos, CA, USA: IEEE Computer Society, May 2024, pp. 129--139. [Online]. Available: \url{https://doi.ieeecomputersociety.org/10.1109/ICST60714.2024.00020}
\BIBentrySTDinterwordspacing

\end{thebibliography}
\end{document}